\documentclass[aoas,MSNbibl,nameyear,seceqn,dvips]{arximspdf}
\usepackage{dcolumn}
\usepackage{graphicx}

\doi{10.1214/10-AOAS405}
\volume{5}
\issue{2A}
\pubyear{2011}
\firstpage{773}
\lastpage{797}

\makeatletter
\newcolumntype{d}[1]{D{.}{.}{#1}}
\newcommand{\notag}{\nonumber}
\renewcommand{\epsilon}{\varepsilon}
\newcommand{\overset}{\stackrel}

\makeatother

\begin{document}
\begin{frontmatter}

\title{Missing data in value-added modeling of teacher effects\thanksref{T1}}
\runtitle{Missing data in value-added models}

\begin{aug}
\author[a]{\fnms{Daniel F.} \snm{McCaffrey}\corref{}\ead[label=e1]{danielm@rand.org}}
\and
\author[a]{\fnms{J. R.} \snm{Lockwood}\ead[label=e2]{lockwood@rand.org}}
\thankstext{T1}{This material is based on work supported by the US Department
of Education Institute of Education Sciences under Grant Nos
R305U040005 and R305D090011,
and the RAND Corporation. Any opinions, findings and conclusions or
recommendations expressed in this material are those of the author(s) and
do not necessarily reflect the views of these organizations.}
\runauthor{D. F. McCaffrey and J. R. Lockwood}
\affiliation{The RAND Corporation}
\address[a]{The RAND Corporation\\
4570 Fifth Avenue, Suite 600 \\
Pittsburgh, Pennsylvania 15213\\
USA\\
\printead{e1}\\
\hphantom{E-mail: }\printead*{e2}} %adresu isvedimo komanda gale!
\end{aug}

% HISTORY:
\received{\smonth{1} \syear{2009}}
\revised{\smonth{7} \syear{2010}}

% ABSTRACT
%
\begin{abstract}
The increasing availability of longitudinal student achievement data has
heightened interest among researchers, educators and policy makers in using
these data to evaluate educational inputs, as well as for school and possibly
teacher accountability. Researchers have developed elaborate
``value-added models'' of these longitudinal data to estimate the
effects of
educational inputs (e.g., teachers or schools) on student achievement while
using prior achievement to adjust for nonrandom assignment of students to
schools and classes. A challenge to such modeling efforts is the extensive
numbers of students with incomplete records and the tendency for those
students to be lower achieving. These conditions create the potential for
results to be sensitive to violations of the assumption that data are missing
at random, which is commonly used when estimating model parameters. The
current study extends recent value-added modeling approaches for
longitudinal student achievement data Lockwood et al. [\textit{J.
Educ. Behav. Statist.}
\textbf{32} (2007)
125--150] to allow data to
be missing not at random via random effects selection and pattern mixture
models, and applies those methods to data from a large urban school district
to estimate effects of elementary school mathematics teachers. We find that
allowing the data to be missing not at random has little impact on estimated
teacher effects. The robustness of estimated teacher effects to the
missing data assumptions appears to result from both the relatively small
impact of model specification on estimated student effects compared
with the
large variability in teacher effects and the downweighting of scores from
students with incomplete data.
\end{abstract}

% KEYWORDS
%

\end{frontmatter}
\setcounter{footnote}{1}
\newpage

%s1 ###
\section{Introduction}\label{intro}

%s1.1 ###
\subsection{Introduction to value-added modeling}

Over the last several years testing of students with standardized
achievement assessments has increased dramatically. As a consequence
of the federal No Child Left Behind Act, nearly all public school
students in the United States are tested in reading and mathematics
in grades 3--8 and one grade in high school, with additional
testing in science. Again spurred by federal policy, states and
individual school districts are linking %\vadjust{\goodbreak}
the scores for students over
time to create longitudinal achievement databases. The data
typically include students' annual total raw or scale scores on the
state accountability tests in English language arts or reading and
mathematics, without individual item scores. Less frequently the
data also include science and social studies scores. Additional
administrative data from the school districts or states are required
to link student scores to the teachers who provided instruction. Due
to greater data availability, longitudinal data analysis is now a
common practice in research on identifying effective teaching
practices, measuring the impacts of teacher credentialing and
training, and evaluating other educational interventions
[\citet{BifuLadd2004};
\citet{GoldhaberCTQ2004};
\citet{HanuKainRivk2002};
\citet{HarrSass2006b};
\citet{LeMosaic2006};
\citet{SchaThum2004};
\citet{ZimmerCSO2003}].
Recent computational advances and empirical findings about the
impacts of individual teachers have also intensified interest in
``value-added'' methods (VAM), where the trajectories of students'
test scores are used to estimate the contributions of individual
teachers or schools to student achievement
[\citet{BallSandWrig2004};
\citet{Braun2005a};
\citet{JacoLefg2006};
\citet{KaneRockStai2006};
\citet{LissitzVAM2005};
\citet{McCaLockKore2003};
\citet{SandSaxtHorn1997}].
The basic notion of VAM is to use longitudinal test score data to
adjust for nonrandom assignment of students to schools and classes
when estimating the effects of educational inputs on achievement.
%% JRL DELETED: With teacher and school accountability at the forefront
%of education
%% policy \edit{\cite{FedReg:2009}}, and with educators and researchers
%% seeking more sophisticated ways of putting test score data to good
%% use, longitudinal methods are likely to remain critical to education
%% research.
%Fed Register is at
%%http://www.ed.gov/legislation/FedRegister/proprule/2009-3/072909d.pdf

%s1.2 ###
\subsection{Missing test score data in value-added modeling}

Longitudinal test score data commonly are incomplete for a large
percentage of the students represented in any given data set. For
instance, across data sets
from several large school systems, we found that anywhere from about
42 to nearly 80 percent of students were missing data from at least
one year out of four or five years of testing. The sequential
multi-membership models used by statisticians for the longitudinal
test score data
[\citet{RaudBryk2002};
\citet{McCaLockKoreLouiHami2004};
\citet{LockMcCaMariSeto2007}]
assume that incomplete data are missing at random [MAR,
\citet{LittRubi2002}]. MAR requires that, conditional on the
observed data, the unobserved scores for students with incomplete
data have the same distribution as the corresponding scores from
students for whom they are observed. In other words, the probability
that data are observed depends only on the observed data in the
model and not on unobserved achievement scores or latent variables
describing students' general level of achievement.

As noted in Singer and Willet (\citeyear{SingWill2003}), the tenability
of missing data assumptions should not be taken for granted, but
rather should be investigated to the extent possible. Such
explorations of the MAR assumption seem particularly important for
value-added modeling given that the proportion of incomplete records
is high, the VA estimates are proposed for high stakes decisions
(e.g., teacher tenure and pay), and the sources of missing data
include the following: students who failed to take a test in a given
year due to
extensive absenteeism, refused to complete the exam, or cheated; the
exclusion of students with disabilities or limited English language
proficiency from testing or testing them with distinct forms
yielding scores not comparable to those of other students; exclusion
of scores after a student is retained in grade because the
grade-level of testing differs from the remainder of the cohort; and
student transfer. Many students transfer schools, especially in
urban and rural districts [\citet{GAO1994}] and school district
administrative data systems typically cannot track students who
transfer from the district. Consequently, annual transfers into and
out of the educational agency of interest each year create data with
dropout, drop-in and intermittently missing scores. Even statewide
databases can have large numbers of students dropping into and out
of the systems as students transfer among states, in and out of
private schools, or from foreign countries.

As a result of the sources of missing data, incomplete test scores
are associated with lower achievement because students with
disabilities and those retained in a grade are generally
lower-achieving, as are students who are habitually absent
[\citet{DunnKadaGarr2003}] and highly mobile
[\citet{HanuKainRivk2004b};
\citet{MehaReyn2004};
\citet{Rumb2003};
\citet{StraDemi2006};
\citet{GAO1994}]. Students with incomplete data might differ from other
students even after controlling for their observed scores.
Measurement error in the tests means that conditioning on observed
test scores might fail to account for differences between the
achievement of students with and without observed test scores.
Similarly, test scores are influenced by multiple historical factors
with potentially different contributions to achievement, and
observed scores may not accurately capture all these factors and
their differences between students with complete and incomplete
data. For instance, highly mobile students differ in many ways from
other students, including greater incidence of emotional and
behavioral problems, and poorer health outcomes, even after
controlling
for other risk factors such as demographic variables [\citet
{WoodETAL2003};
\citet{SimpFowl1994};
\citet{ElliMcGu2000}]. %

However, the literature provides no thorough empirical
investigations of the pivotal MAR assumption, even though incomplete
data are widely discussed as a potential source of bias in estimated
teacher effects and thus a~potential threat to the utility of
value-added models [\citet{Braun2005b};
\citet{McCaLockKore2003};
\citet{Kupe2003}].
A few authors [\citet{Wright2004};
\citet{McCaLockMariSeto2005}] have
considered the implications of violations of MAR for estimating
teacher effects through simulation studies. In these studies, data
were generated and then deleted according to various scenarios,
including those where data were missing not at random (MNAR), and
then used to estimate teacher effects. Generally, these studies have
found that estimates of school or teacher effects produced by random
effects models used for VAM are robust to violations of the MAR
assumptions and do not show appreciable bias except when the
probability that scores are observed is very strongly correlated
with the student achievement or growth in achievement. However,
these studies did not consider the implications of relaxing the MAR
assumption on estimated teacher effects, and there are no examples
in the value-added literature in which models that allow data to be
MNAR are fit to real student test score data.

%{\bf JRL: FEELS LIKE WE SHOULD COMBINE FOLLOWING PARAGRAPH WITH LAST
%PARAGRAPH IN SECTION
%1.3. BELONGS THERE LOGICALLY AND WOULD ALSO AVOID THE FORWARD
%REFERENCE OF
%LITTLE'S TERMINOLOGY.}
%investigation. Following the suggestions of Hedeker and Gibbons
%assumption
%by considering two alternative MNAR model specifications: selection
%and a
%pattern mixture models (described below), and compare these to a
%cross-classified multi-membership hierarchical model assuming MAR.
%MNAR models
%have not previously been used with these hierarchical models where the
%cluster
%variables (i.e., teacher effects) are of interest.}
%

%s1.3 ###
\subsection{MNAR models}

The statistics literature has seen the development and application
of numerous models for MNAR data. Many of these models apply to
longitudinal data in which participants drop out of the study, and
time until dropout is modeled simultaneously with the outcome data
of interest
[\citet{GuoCarl2004};
\citet{TenhReboMillKuns2002};
\citet{WuCarr1988}]. Others
allow the probability of dropout to depend directly on the observed
and unobserved outcomes [\citet{DiggKenw1994}]. Little
(\citeyear{Litt1995}) provides two general classes of models for MNAR
data: selection models, in which the probability of data being
observed is modeled conditional on the observed data, and pattern
mixture models, in which the joint distribution of longitudinal data
and missing data indicators is partitioned by response pattern so
that the distribution of the longitudinal data (observed and
unobserved) depends on the pattern of responses. Little
(\citeyear{Litt1995}) also develops a selection model in which the
response probability depends on latent effects from the outcome data
models, and several authors have used these models for incomplete
longitudinal data in health applications
[\citet{FollWu1995};
\citet{IbraChenLips2001};
\citet{HedeGibb2006}], and modeling
psychological and attitude scales and item response theory
applications in which individual items that contribute to a~scale or
test score are available for analysis
[\citet{OMuiMous1999};
\citet{MousKnot2003};
\citet{HolmGlas2005};
\citet{KoroETAL2008}].
Pattern mixture models have also been suggested by various authors
for applications in health
[\citet{FitzLairShne2001};
\citet{HedeGibb1997};
\citet{Litt1993}].

Although these models are well established in the statistics
literature, their use in education applications has been limited
primarily to the context of psychological scales and item response
models rather than longitudinal student achievement data like those
used in value-added models. In particular, the MNAR models have not
been adapted to sequential multi-membership models used in VAM,
where the primary focus is on random effects for teachers (or
schools), and not on the individual students or in the fixed effects
which typically are the focus of other applications of MNAR models.
Moreover, in many VAM applications, including the one presented
here, when students are missing a score they also tend to be missing
a link to a teacher because they transferred out of the education
agency of interest and are not being taught by a teacher in the
population of interest. Again, this situation is somewhat unique to
the setting of VAM and its implications for the estimation of the
teacher or school effects is unclear.

Following the suggestions of Hedeker and Gibbons
(\citeyear{HedeGibb2006}) and Singer and Willet (\citeyear{SingWill2003}),
this paper applies two alternative MNAR model specifications: random
effects selection and a pattern mixture model to extend recent
value-added modeling approaches for longitudinal student achievement
data [\citet{LockMcCaMariSeto2007}] to allow data to be missing not
at random. We use these models to estimate teacher effects using a
data set from a large urban school district in which nearly 80
percent of students have incomplete data and compare the MNAR and
MAR specifications. We find that even though the MNAR models better
fit the data, teacher effect estimates from the MNAR and MAR models
are very similar. We then probe for possible explanations for this
similarity.

%s2 ###
\section{Data description}

The data contain mathematics scores on a norm-referenced standardized test
(in which test-takers are scored relative to a~fixed reference population)
for spring testing in 1998--2002 for all students in grades 1--5 in a
large urban US school district. The data are ``vertically linked,''
meaning that the test scores are on a common scale across grades, so that
growth in achievement from one grade to the next can be measured. For our
analyses we standardized the test scores by subtracting 400 and
dividing by
40. We did this to make the variances approximately one and to keep the
scores positive with a mean that was consistent with the scale of the
variance. Although this rescaling had no effect on our results, it
facilitated some computations and interpretations of results.

For this analysis, we focused on estimating effects on mathematics
achievement for teachers of grade 1 during the 1997--1998 school year,
grade 2 during the 1998--1999 school year, grade 3 during the 1999--2000
school year, grade 4 during the 2000--2001 school year and grade 5
during the 2001--2002 school year. A total of 10,332 students in our
data link to these teachers.\footnote{Students were linked to the
teachers who administered the tests. These teachers might not always
be the teachers who provided instruction but for elementary schools
they typically are.} However, for some of these students the data
include no valid test scores or had other problems such as unusual
patterns of grades across years that suggested incorrect linking of
student records or other errors. We deleted records for these
students. The final data set includes 9,295 students with 31 unique
observation patterns (patterns of missing and observed test scores
over time). The data are available in the supplemental materials
[\citet{McCaLockAOAS405Supp}].

\begin{figure}

\includegraphics{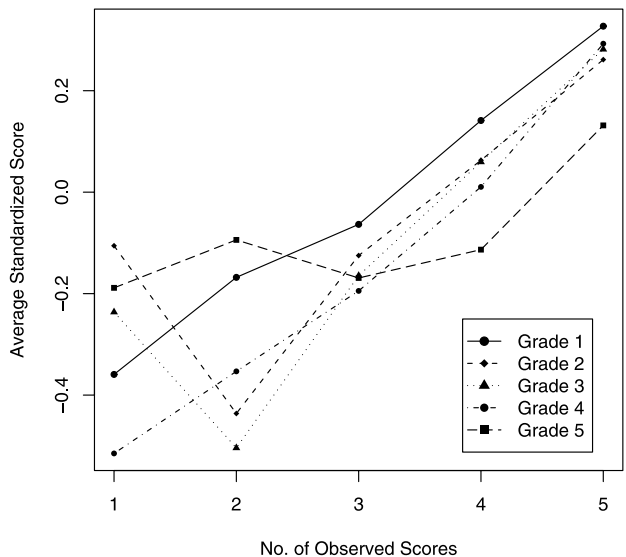}

\caption{Standardized score means by grade of testing
as a function of a student's number of observed scores.}
\label{fig:MeanNobs}
\end{figure}

Missing data are extremely common for the students in our sample.
Overall, only about 21 percent of the students have fully observed
scores, while 29, 20, 16 and 14 percent have one to four observed
scores, respectively. Consistent with previous research, students
with fewer scores tend to be lower-scoring. As shown in
Figure~\ref{fig:MeanNobs}, students with five observed scores on
average are often scoring more than half a standard deviation higher
than students with one or two observed scores.

%f1 ###

Moreover, the distribution across teachers of students with
differing numbers of observed scores is not balanced. Across
teachers, the proportion of students with complete test scores
averages about 37 percent\footnote{The average percentage of
students with
complete scores at the teacher level exceeds the marginal percentage of
students with complete data because in each year, only students linked to
teachers in that year are used to calculate the percentages, and missing
test scores are nearly always associated with a missing teacher link in
these data.} but ranges anywhere from 0 to 100 percent in every grade.
Consequently, violation of the MAR assumption is unlikely to have an
equal effect on all teachers and could lead to differential bias in
estimated teacher effects. %% see
%dc-percentage-students-with-complete-data-by-tch.R for gettiing
%percents
%% \begin{figure}
%%   \caption[]{Distribution of percentage of students in
%each class with
%% five years of mathematics scores by grade.} \label{fig:PcntCmplt}
%% %
%% \end{figure}

%s3 ###
\section{Models}

Several authors [\citet{SandSaxtHorn1997};
\citet{McCaLockKoreLouiHami2004};
\citet{LockMcCaMariSeto2007};
\citet{RaudBryk2002}] have proposed random effects
models for analyzing longitudinal student test score data, with
scores correlated within students over time and across students
sharing either current or past teachers. Lockwood et al.
(\citeyear{LockMcCaMariSeto2007}) applied the following model to our
test score data to estimate random
effects for classroom membership:
%
%e3.1 ###
\begin{eqnarray}\label{eq:mar1}
Y_{it} & =&\mu_{t} + \sum_{t^{*} \leq t} \alpha_{tt^{*}} \bolds{\phi
}_{it^{*}}'\bolds{\theta}_{t^{*}} + \delta_i +
\epsilon_{it},\notag\\
\bolds{\theta}_{t^{*}} & =&
(\theta_{t^{*}1},\ldots ,\theta_{t^{*}J_{t^{*}}})',  \qquad  \theta
_{t^{*}j}{\overset{\mathrm{i.i.d.}}{\sim}}
N(0, \tau^2_{t^{*}}), \\
\delta_i
&{\overset{\mathrm{i.i.d.}}{\sim}}&N(0,\nu^2),  \qquad  \epsilon_{it}{\overset
{\mathrm{i.i.d.}}{\sim}}
N(0, \sigma_t^2).\notag
\end{eqnarray}
The test score $Y_{it}$ for student $i$ in year $t$, $t=1,\ldots ,5$,
depend on $\mu_t$, the annual mean, as well as random effects
$\bolds{\theta}_t$ for classroom membership for each year. The vectors
$\bolds{\phi}_{it}$, with $\phi_{itj}$ equal to one if student $i$ was
taught by teacher $j$ in year $t$ and zero otherwise, link students
to their classroom memberships. In many VAM applications, these
classroom effects are treated as ``teacher effects,'' and we use
that term for consistency with the literature and for simplicity in
presentation. However, the variability in scores at the classroom
level may reflect teacher performance as well as other potential
sources such as schooling and community inputs, peers and omitted
individual student-level characteristics
[\citeauthor{McCaLockKoreLouiHami2004} (\citeyear{McCaLockKore2003,McCaLockKoreLouiHami2004})].

Model (\ref{eq:mar1}) includes terms for students' current and prior
classroom assignments with prior assignments weighted by the
$\alpha_{tt^{*}}$, allowing correlation among scores for students
who shared a classroom in the past, that can change over time by
amounts that are determined by the data. By definition,
$\alpha_{tt^{*}}=1$ for $t^{*} = t$. Because student classroom
assignments change annually, each student is a member of multiple
cluster units from which scores might be correlated. The model is
thus called a multi-membership model [\citet{BrowGoldRasb2001}] and
because the different memberships occur sequentially rather than
simultaneously, we refer to the model as a sequential
multi-membership model.

The $\delta_i$ are random student effects.
%which we refer to
%$\delta_i$ as the student's ``general level of achievement'' because
%it is persistent across the repeated achievement measures and
%determines the student's average achievement relative to the annual
%means for all students.
McCaffrey et al.
(\citeyear{McCaLockKoreLouiHami2004}) and Lockwood et
al. (\citeyear{LockMcCaMariSeto2007}) consider a more general model in
which the residual error terms are assumed to be multivariate normal
with mean vector $\mathbf{0}$ and an unstructured variance--covariance
matrix. Our specification of $(\delta_i + \epsilon_{it})$ for the
error terms is consistent with random effects models considered by
other authors [\citet{RaudBryk2002}] and supports generalization to
our MNAR models.
%
%$\bolds{\phi}_{it^{*}}'\bolds{\theta}_{t^{*}}$, $t^{*} < t$, are unknown
%because the student was not enrolled in the district or retained in
%grade and not taught by a teacher in the sample. However, these
%unobserved values are needed for modeling $Y_{it}$ via
%Model \ref{eq:mar1}. Lockwood et al. \cite{LockMcCaMariSeto2007}
%considered two alternative approaches to this problem in their study
%of these data. Their first approach assigned a student a unique
%random teacher effect from a population distinct from the population
%of interest every time a student's prior teacher was unknown. Their
%second approach assumed that unknown prior teachers had zero effect.
%Lockwood and colleagues found that estimated teacher effects were
%robust to these two approaches \cite{LockMcCaMariSeto2007}. They
%thus focused on the second, simpler approach, and we do the same.}
%
%{\bf JRL: COULD WE REPLACE PREVIOUS PARAGRAPH WITH SHORTER VERSION: ``

When students drop into the sample at time $t$, the identities of
their teachers prior to time $t$ are unknown, yet are required for
modeling $Y_{it}$ via Model~(\ref{eq:mar1}). Lockwood et al.
(\citeyear{LockMcCaMariSeto2007}) demonstrated that estimated teacher
effects were robust to different approaches for handling this
problem, including a~simple approach that assumes that unknown prior
teachers have zero effect, and we use that approach here.

Following Lockwood et al. (\citeyear{LockMcCaMariSeto2007}), we fit
Model (\ref{eq:mar1}) to the incomplete mathematics test score data
described above using a Bayesian approach with relatively
noninformative priors via data augmentation that treated the
unobserved scores as MAR. We refer to this as our MAR model. We
then modify Model (\ref{eq:mar1}) to consider MNAR models for the
unobserved achievement scores. In the terminology of Little
(\citeyear{Litt1995}), the expanded models include random effects
selection models and a pattern mixture model.

%s3.1 ###
\subsection{Selection model}
The selection model makes the following additional assumption to
Model (\ref{eq:mar1}):

\begin{enumerate}
\item
$\operatorname{Pr}(n_{i} \leq k) = \frac{e^{a_k + \beta\delta_i}}{1+ e^{a_k +
\beta\delta_i}}$, where $n_{i}=1, \ldots , 5,$ equals the number of
observed mathematics test scores for student $i$.
\end{enumerate}

Assumption 1 states that the number of observed scores $n_i$ depends on the
unobserved student effect $\delta_i$. Students who would tend to score high
relative to the mean have a different probability of being observed
each year
than students who would generally tend to score lower. This is a
plausible model
for selection given that mobility and grade retention are the most common
sources of incomplete data, and, as noted previously, these
characteristics are
associated with lower achievement. The model is MNAR because the probability
that a score is observed depends on the latent student effect, not on observed
scores.
%% and students with the same observed scores can have different
%probabilities of
%% providing a score for each year of testing.
We use the notation ``SEL'' to
refer to estimates from this model to distinguish them from the other models.

Because $n_i$ depends on $\delta$, by Bayes' rule the distribution of
$\delta$
conditional on $n_i$ is a function of $n_i$. Consequently, assumption 1
implicitly makes $n_i$ a~predictor of student achievement. The model, therefore,
provides a means of using the number of observed scores to inform the prediction
of observed achievement scores, which influences the adjustments for student
sorting into classes and ultimately the estimates of teacher effects.

As discussed in Hedeker and Gibbons (\citeyear{HedeGibb2006}), the space
of MNAR models is very large and any sensitivity analysis of missing
data assumptions should consider multiple models. Per that advice,
we considered the following alternative selection model. Let
$r_{it}$ equal one if student $i$ has an observed score in year
$t=1, \ldots , 5$ and zero otherwise. The alternative selection model
replaces assumption 1 with assumption 1a.
\begin{enumerate}
\item[1a.] Conditional on $\delta_i$, $r_{it}$ are independent with
$\operatorname{Pr}(r_{it} = 1 |\delta_i$) = $\frac{e^{a_t +
\beta_t\delta_i}}{1+e^{a_t +
\beta_t\delta_i}}$.
\end{enumerate}
Otherwise the models are the same. This model is similar to those
considered by other authors for modeling item nonresponse in
attitude surveys and multi-item tests
[\citet{OMuiMous1999};
\citet{MousKnot2003};
\citet{HolmGlas2005};
\citet{KoroETAL2008}],
although those models also sometimes include a latent response
propensity variable.

%s3.2 ###
\subsection{Pattern mixture model}

Let $\mathbf{r}_i=(r_{i1},\ldots ,r_{i5})'$, the student's pattern of
responses. Given that there are five years of testing and every
student has at least one observed score, $\mathbf{r}_i$ equals
$\mathbf{r}^k$, for $k=1,\ldots ,31$ possible response patterns. The
pattern mixture model makes the following assumption to extend
Model (\ref{eq:mar1}):
\begin{enumerate}
\item[2.] Given $\mathbf{r}_i = \mathbf{r}^k$,
%
%e3.2 ###
\begin{eqnarray}\label{eq:patmix1}
Y_{it} & =&\mu_{kt} + \sum_{t^{*} \leq t} \alpha_{tt^{*}}
\bolds{\phi}_{it^{*}}'\bolds{\theta}_{t^{*}} + \delta_{i} + \zeta_{it},
\notag\\
\delta_{i} & {\overset{\mathrm{i.i.d.}}{\sim}}&N(0, \nu_{k}^2), \qquad          \zeta
_{it} {\overset{\mathrm{i.i.d.}}{\sim}} N(0,
\sigma_{kt}^2),\\
\theta_{tj} & {\overset{\mathrm{i.i.d.}}{\sim}}& N(0, \tau^2_t).\notag
\end{eqnarray}
\end{enumerate}
We only estimate parameters for $t$'s corresponding to the observed
years of data for students with pattern $k$. By assumption 2,
teacher effects and the out-year weights for those effects
($\alpha_{tt*}, t*<t$) do not depend on the student's response
pattern. We use ``PMIX'' to refer to this model.

Although all 31 possible response patterns appear in our data, each
of five patterns occurs for less than 10 students and one pattern
occurs for just 20 students. We combined these six patterns into a
single group with common annual means and variance components
regardless of the specific response pattern for a student in this
group. Hence, we fit 25 different sets of mean and variance
parameters corresponding to different response patterns or groups of
patterns. Combining these rare patterns was a pragmatic choice to
avoid overfitting with very small samples. Given how rare and
dispersed students with these patterns were, we did not think
misspecification would yield significant bias to any individual
teacher. We ran models without these students and even greater
combining of patterns and had similar results. For each of the five
patterns in which the students had a single observed score, we
estimated the variance of $\delta_{ki}+\zeta_{kit}$ without
specifying student effects or separate variance components for the
student effects and annual residuals.

%% \edit{Should we remove this line of argument throughout?}The pattern
%% mixture model explicitly allows the mean achievement to depend on
%% the pattern of missing data which is analogous to including these
%% variables as independent variables in a model for the mean. This
%% could result in overcorrecting for pattern of missing scores if true
%% teacher (or classroom) effects are correlated with the proportion of
%% students with a given response pattern across classes. The pattern
%% mixture model makes an explicit adjustment for response pattern and
%% the selection model makes an implicit adjustment for pattern (in
%% terms of number of observed scores), which might result in
%% differences for the estimated teacher effects from these two models.

%s3.3 ###
\subsection{Prior distributions and estimation}

Following the work of Lockwood et al.
(\citeyear{LockMcCaMariSeto2007}), we estimated the models using a
Bayesian approach with priors chosen to be relatively uninformative:
$\mu_t$ or $\mu_{tk}$ are independent $N(0, 10^6)$, $t=1,\ldots ,5$,
$k=1,\ldots ,25$; $\alpha_{tt^{*}} \sim N(0,10^6)$, $t=1,\ldots ,5$,\vspace*{-2pt}
$t^{*}=1,\ldots ,t$; $\theta_{tj}{\overset{\mathrm{i.i.d.}}{\sim}}N(0,\tau^2_t)$,
$j=1,\ldots ,J_t$, $\tau_t$, $t=1, \ldots , 5$, are $\operatorname{uniform}(0,0.7)$,
$\delta_i {\overset{\mathrm{i.i.d.}}{\sim}} N(0, \nu^2)$, $\nu$ is $\operatorname{uniform}(0,2)$,
and $\sigma_t$'s are $\operatorname{uniform}(0,1)$. For the selection model,
SEL, the parameters for the models for number of responses ($a$,
$\beta$) are independent $N(0, 100)$ variables. For the alternative
selection model the $a_t$'s and $\beta_t$'s are $N(0,10)$ variables.
All parameters are independent of other parameters in the model and
all hyperparameters are independent of other hyperparameters.

We implemented the models in WinBUGS [\citet{winbugs-new}]. WinBUGS
code used for fitting   all   models reported in
this article can be found in the supplement
[\citet{McCaLockAOAS405Supp}]. For each model, we ``burned in''
three independent chains each for 5000 iterations and based our
inferences on 5000 post-burn-in iterations. We diagnosed
convergence of the chains using the Gelman--Rubin diagnostic
[\citet{GelmRubi1992}] implemented in the \texttt{coda} package
[\citet{BestCowlVine1995}] for the R statistics environment
[\citet{RDCT}]. The 5000 burn-in iterations were clearly sufficient
for convergence of model parameters. Across all the parameters
including teacher effects and student effects (in the selection
models), the Gelman--Rubin statistics were generally very close to one
and always less than 1.05.

%t1 ###
\begin{table}[h!]
\tabcolsep=0pt
\caption{Posterior means and standard deviations for parameters
other than teacher and student effects from MAR and SEL
models}\label{tab:parm1}
\begin{tabular*}{\textwidth}{@{\extracolsep{\fill}}lccd{2.2}c@{}}
\hline
& \multicolumn{2}{c}{\textbf{MAR}} & \multicolumn{2}{c@{}}{\textbf{SEL}}\\[-5pt]
& \multicolumn{2}{c}{\hrulefill} & \multicolumn{2}{c@{}}{\hrulefill}\\
& \textbf{Posterior} & \textbf{Posterior} & \multicolumn{1}{c}{\textbf{Posterior}} & \textbf{Posterior} \\
\textbf{Parameter} & \textbf{mean} & \textbf{std. dev.} & \multicolumn{1}{c}{\textbf{mean}} & \textbf{std. dev.}\\
\hline
$\mu_1$ & 3.39 & 0.03 & 3.44 & 0.03 \\
$\mu_2$ & 3.98 & 0.03 & 4.01 & 0.03 \\
$\mu_3$ & 4.70 & 0.03 & 4.69 & 0.02 \\
$\mu_4$ & 5.29 & 0.02 & 5.26 & 0.02 \\
$\mu_5$ & 6.00 & 0.03 & 5.96 & 0.03 \\
$\tau_1$ & 0.65 & 0.03 & 0.63 & 0.03 \\
$\tau_2$ & 0.57 & 0.03 & 0.56 & 0.03 \\
$\tau_3$ & 0.55 & 0.03 & 0.54 & 0.03 \\
$\tau_4$ & 0.43 & 0.02 & 0.42 & 0.02 \\
$\tau_5$ & 0.42 & 0.02 & 0.42 & 0.02 \\
$\alpha_{21}$ & 0.16 & 0.02 & 0.14 & 0.03 \\
$\alpha_{31}$ & 0.15 & 0.02 & 0.13 & 0.03 \\
$\alpha_{32}$ & 0.20 & 0.02 & 0.19 & 0.02 \\
$\alpha_{41}$ & 0.12 & 0.02 & 0.09 & 0.02 \\
$\alpha_{42}$ & 0.11 & 0.02 & 0.10 & 0.02 \\
$\alpha_{43}$ & 0.14 & 0.02 & 0.11 & 0.02 \\
$\alpha_{51}$ & 0.11 & 0.02 & 0.08 & 0.03 \\
$\alpha_{52}$ & 0.14 & 0.02 & 0.13 & 0.02 \\
$\alpha_{53}$ & 0.09 & 0.02 & 0.06 & 0.02 \\
$\alpha_{54}$ & 0.34 & 0.03 & 0.34 & 0.03 \\
$\nu$ & 0.71 & 0.01 & 0.73 & 0.01 \\
$\sigma_1$ & 0.58 & 0.01 & 0.57 & 0.01 \\
$\sigma_2$ & 0.47 & 0.01 & 0.47 & 0.01 \\
$\sigma_3$ & 0.45 & 0.01 & 0.45 & 0.01 \\
$\sigma_4$ & 0.37 & 0.01 & 0.37 & 0.01 \\
$\sigma_5$ & 0.37 & 0.01 & 0.37 & 0.01 \\
$a_1$ & NA & NA & -1.00 & 0.02 \\
$a_2$ & NA & NA & 0.90 & 0.02 \\
$a_3$ & NA & NA & 0.71 & 0.02 \\
$a_4$ & NA & NA & 0.79 & 0.02 \\
$\beta$ & NA & NA & -0.83 & 0.03\\
\hline
\end{tabular*}
\end{table}

%s4 ###
\section{Results}

%s4.1 ###
\subsection{Selection models}
The estimate of the model parameters for MAR and SEL other than
teacher and student effects are presented in Table \ref{tab:parm1} of
the \hyperref[appm]{Appendix}. The selection model found that the number of
observed scores is related to students' unobserved general levels of
achievement $\delta_i$. The posterior mean and standard deviation
for $\beta$ were $-$0.83 and 0.03, respectively. At the mean for
$\beta$, a~student with an effect of $\delta=0.72$ (one standard
deviation above the prior mean of zero) would have a probability of 0.31
of completing all five years of testing, whereas the probability
for a~student with an effect of $\delta=- 0.72$ (one standard
deviation below the mean) would be only  0.12.

\begin{figure}

\includegraphics{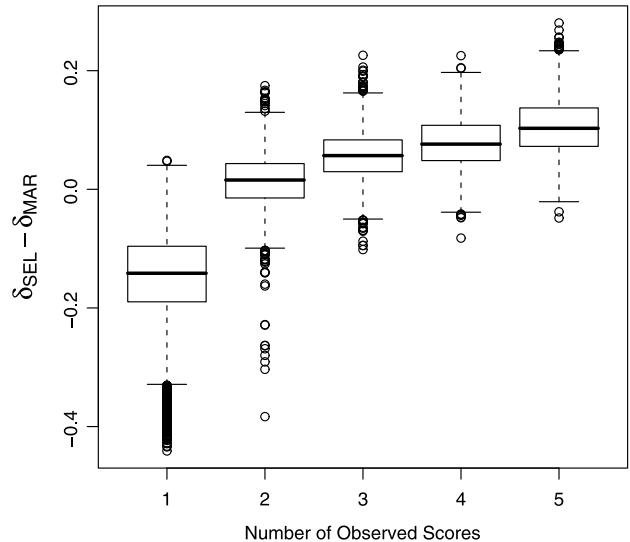}%
\vspace*{-5pt}
\caption{Distributions of differences in the posterior means for each
student effect from the selection model ($\delta_{\mathrm{SEL}}$) and the MAR
model ($\delta_{\mathrm{MAR}}$). All effects are standardized by the posterior
means for their respective
standard deviations ($\nu$). Distributions are presented by the number
of observed mathematics scores.}
\label{fig:deltas}
\vspace*{-5pt}
\end{figure}

 Figure \ref{fig:deltas} shows the effect that modeling
the number of observed scores has on estimated student effects. We
estimated each student's effect using the posterior mean from the selection
model ($\delta_{\mathrm{SEL}}$) and we also estimated it using the posterior mean
from Model (\ref{eq:mar1}) assuming MAR ($\delta_{\mathrm{MAR}}$). For each student
we calculated the difference in the two alternative estimates of his or her
effect ($\delta_{\mathrm{SEL}} - \delta_{\mathrm{MAR}}$) where the estimates were
standardized by the corresponding posterior mean for the standard deviation
in student effects. The left panel of Figure \ref{fig:deltas} plots the
distribution of these differences by the number of observed
scores.
%
%
% \caption[]{Distributions of differences in the posterior means for
%each
% student effect from the selection model ($\delta_{SEL}$) and the MAR
% model ($\delta_{MAR}$) (left panel) and the pattern mixture model
% ($\delta_{PMIX}$) and the MAR model ($\delta_{MAR}$) (right panel).
% All effects are standardized the posterior means for their respective
% standard deviations ($\nu$). Distributions are presented by the number
% of observed mathematics scores. For the PMIX model, distributions are
% shown only for students with two or more observed scores because this
% model did not explicitly estimate student effects for the response
% patterns for students with a single observed score.}
% \label{fig:deltas}
% \begin{tabular}{lr}
%   %
%&
% %
% \end{tabular}

%f2 ###

The figure clearly shows that modeling the number of observed scores
provides additional information in estimating each student's effect,
and, as would be expected, the richer model generally leads to
increases in the estimates for students with many observed scores
and decreases in the estimates for students with few observed
scores. Although modeling the number of test scores provides
additional information about the mean of each student's effect, it
does not significantly reduce uncertainty about the student effects.
Across all students the posterior standard deviation of the student
effect from SEL is 99 percent as large as the\ corresponding
posterior standard deviation from the MAR model and the relative
sizes of the posterior standard deviations do not depend on the
number of observed scores.

%f3 ###
\begin{figure}

\includegraphics{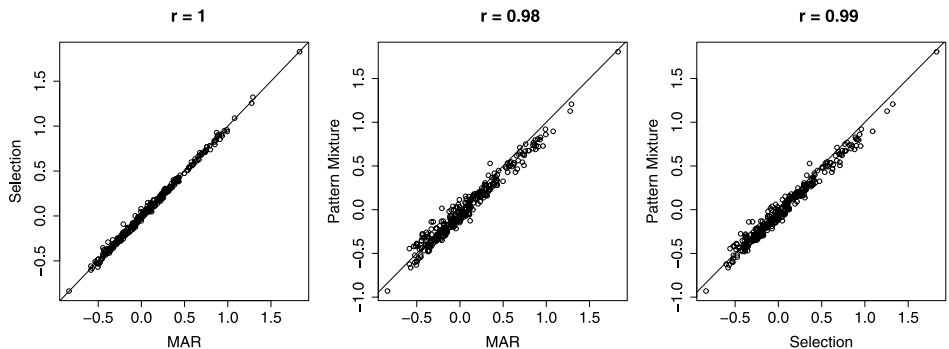}%
\vspace*{-5pt}
\caption{Scatter plots of posterior means for fourth
grade teacher effects from selection, pattern mixture and MAR
models.} \label{fig:thetas}\vspace*{-5pt}
\end{figure}

We used the Deviance Information Criterion [DIC;
\citet{SpieBestCarlLind2002}] as calculated in WinBUGS to compare
the fits of the MAR and the selection model. DIC is a model
comparison criterion for Bayesian models that combines a measure of
model fit and model complexity to indicate which, among a set of
models being compared, is preferred (as indicated by the smallest
DIC value). Apart from a normalizing constant that depends on only
the observed data and thus does not affect model comparison, DIC is
given by $-4\bar{L}+2L(\bar{\bolds{\omega}})$, where $\bar{L}$ is the
posterior mean of the log-likelihood function and
$L(\bar{\bolds{\omega}})$ is the log-likelihood function evaluated at
the posterior mean $\bar{\bolds{\omega}}$ of the model parameters. We
obtained DIC values of 40,824 for the MAR model and 40,658 for the
selection model. As smaller values of DIC indicate preferred models,
with differences of 10 or more DIC points generally considered to be
important, the selection model is clearly preferred to the MAR
alternative.

\begin{figure}

\includegraphics{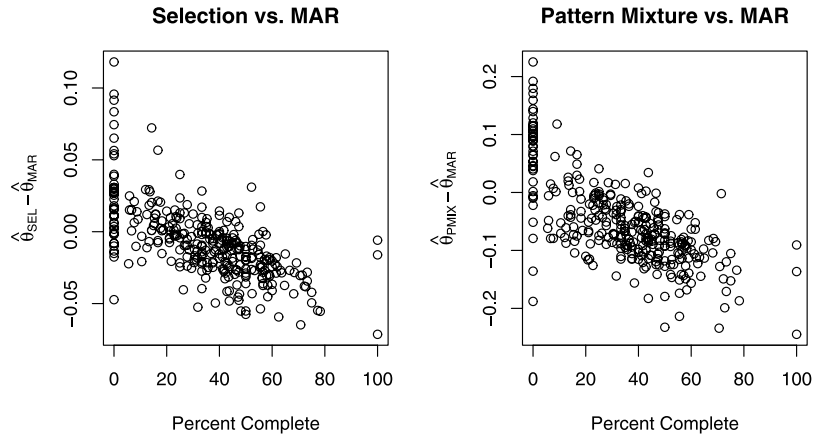}

 \caption{Scatter plots of differences in posterior
means for fourth grade teacher effects from selection and MAR model
\textup{(left panel)} or pattern mixture and MAR model \textup{(right panel)} versus
the proportion of students with five years of test scores.}
\label{fig:dthetas}
\end{figure}

Although the selection model better fits the data and had an impact
on the estimates of individual student effects, it did not have any
notable effect on estimates of teacher effects. The correlation
between estimated effects from the two models was 0.99, 1.00, 1.00,
1.00 and 1.00 for teachers from grade 1 to~5, respectively. The
left panel of Figure \ref{fig:thetas} gives a scatter plot of the
two sets of estimated effects for grade 4 teachers and shows that
two sets of estimates were not only highly correlated but are nearly
identical. Scatter plots for other grades are similar. However, the
small differences that do exist between the estimated teacher
effects from the two models are generally related to the amount of
information available on teachers' students. As shown in the left
panel of Figure \ref{fig:dthetas}, relative to those from the MAR
model, estimated teacher effects from the selection model tended to
decrease with the proportion of students in the classroom with
complete data. This is because student effects for students with
complete data were generally estimated to be higher with the
selection model than with the MAR model and, consequently, these
students' higher than average scores were attributed by the
selection model to the student rather than the teacher, whereas the
MAR model attributed these students' above-average achievement to
their teachers. The differences are generally small because the
differences in the student effects are small (i.e., differences for
individual students in posterior means from the two models account
for about one percent of the overall variance in the student effects
from the MAR model).

%f4 ###

The results from the alternative selection model (assumption~1a) are
nearly identical to those from SEL with estimated teacher effects
from this MNAR model correlated between  0.97 and 1.00 with the
estimate from SEL and almost as highly with the estimates from MAR
(details are in the \hyperref[appm]{Appendix}).

%s4.2 ###
\subsection{Pattern mixture model}

\begin{figure}

\includegraphics{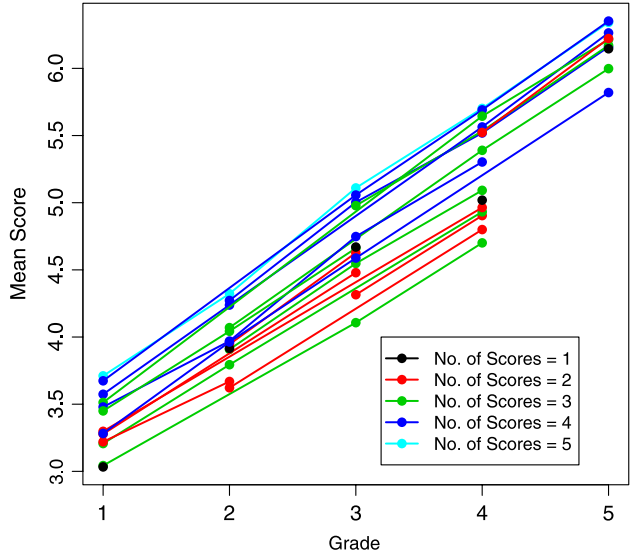}

 \caption{Posterior means for grade specific means
from the pattern mixture model. Means from the same response pattern
are connected by lines and color coded by number of observed scores
for the pattern of response.} \label{fig:pmixmean}
\end{figure}

The results from the pattern mixture models were analogous to those
from the selection model: allowing the data to be MNAR changed our
inferences about student achievement but had very limited effect on
inferences about teachers. Because of differences in the modeling
of student effects, the DIC for the pattern mixture model is not
comparable to the DIC for the other models and we cannot use this
metric to compare models. However, as shown in
Figure \ref{fig:pmixmean} which plots the estimates of the annual
means by pattern, the pattern mixture model clearly demonstrates
that student outcomes differ by response pattern. As expected,
generally, the means are lower for patterns with fewer observed
scores, often by almost a full standard deviation unit. The
differences among patterns are fairly constant across years so that
growth in the mean score across years is relatively similar
regardless of the pattern.

%f5 ###

% intuition about how PMIX works. It's just a shell game of what we are
% calling student and what we are calling pattern. Could we cut this
%whole
% paragraph and corresponding fig and just say in words that the
% adjustments for PMIX come through a combo of pattern means and delta
%and
% the deltas aren't really comparable to those from SEL and MAR?}
The student effects in the pattern mixture model are relative to the
annual pattern means rather than the overall annual means like the
effect  in MAR and SEL models and the effects from PMIX cannot be
directly compared with those of the other models. However, combining
the student effects with the pattern effect yields estimates that
are generally similar to the student effects from MAR.

As with the selection model, the estimated teacher effects from the
pattern mixture and the MAR models were highly correlated and
generally very similar. The center panel of Figure \ref{fig:thetas}
shows close agreement of the PMIX and MAR posterior mean teacher
effects for the grade 4 teacher effects. The correlations between
the two sets of estimates range from 0.98 to 1.00 across grades. The
small differences that do exist are related to the average number of
observed scores for students in the teachers' classes. Again,
because greater numbers of scores result in patterns with generally
higher mean scores, scores for those students are adjusted downward
by the PMIX model relative to the MAR model and teacher effects are
correspondingly adjusted down for teachers with more students with
complete data. The student effects compensate for the adjustment to
the mean, but, as demonstrated for grade 4 teachers in the right
panel of Figure \ref{fig:dthetas}, effects for teachers with
proportionately more students with complete data tend to be somewhat
lower for the PMIX model than the MAR model.

As the high correlations between estimated teacher effects from MAR and the
selection and pattern mixture models would suggest, the estimated teacher
effects from the two alternative MNAR models are also highly correlated
(0.99 or 1.00 for every grade), as demonstrated in the right panel of
Figure \ref{fig:thetas}.

%s5 ###
\section{Discussion}

We applied models allowing data to be missing not at random in a new
context of estimating the effects of classroom assignments using
longitudinal student achievement data and sequential
multi-membership models. We considered both random effects
selection models and a pattern mixture model. Compared with the
existing MAR models, allowing the number or pattern of observed
scores to depend on a student's general level of achievement in the
selection models decreased our estimates of latent effects for
students with very few observed scores and increased our estimates
for students with complete data. The pattern mixture model found
mean achievement was lower for students with the fewest observed
scores and increased across response patterns as a function of the
number of observed scores. Allowing the data to be MNAR changed
teacher effects in the expected directions: compared with the
estimates from the MAR model, estimated teacher effects from the
MNAR models generally decreased with the proportion of students in
the classroom with complete data because the MAR model overestimated
the achievement of students with few scores and underestimated the
achievement of students with many scores.

However, the changes to the estimated teacher effects were generally tiny,
yielding estimates from alternative models that correlate at  0.98 or better,
and inconsequential to inferences about teachers.
% we should check posterior intervals as well to see if we added or lost
% precision}.
This paradoxical finding is likely the result of multiple factors related
to how student test scores contribute to estimated teacher effects.

To understand how student test scores contribute to posterior means
for the teacher effects, we treat the other parameters in the model
as known and consider a general expression for the posterior means.
For a given set of values for the other model parameters, the
teacher effects are given by $\hat{\bolds{\theta}} =
\mathbf{A}\mathbf{R}^{-1}\mathbf{e}$, where $\mathbf{e}$ is a vector of adjusted
scores, $e_{it}=Y_{it}-\mu_{t}$ or $e_{it}=Y_{it}-\mu_{tk}$ for
PMIX, $\mathbf{R}$ is the block diagonal covariance matrix,
($\{\mathbf{R}_i\}$), of the student-level residuals, and $\mathbf{A}$
depends on the inverse of the variance--covariance matrix of the
vector of scores and classroom assignments
[\citet{SearCaseMcCu1992}]. Results on inverse covariance matrices
[\citet{Thei1971}] yield that for student $i$, element $t$ of
$\mathbf{R}_i^{-1}\mathbf{e}_i$ equals the residual from a~regression of
$e_{it}$ on the other $e$ values for the student divided by its
variance. The variance of these residuals declines with the number
of observed scores, as more scores yield a more precise prediction of
$e_{it}$. Consequently, adjusted scores for students with more
complete data get larger weights and have more leverage on estimated
teacher effects than those for students with more missing data.

The differences in weights can be nontrivial. For example, we
calculated~$\mathbf{R}$ using the posterior means of $\nu^2$ and the
$\sigma^2_t$ for the MAR model and compared the resulting weights
for students with differing numbers of observed scores. The weight
given to any adjusted score depends on both the number of observed
scores and the grades in which they were observed.
% because the
%posterior means for $\sigma^2_t$ are  0.34,  0.22,  0.20,  0.13, and  0.13 for
%grades 1 to 5, respectively. Hence, for each possible response
%pattern,
We calculated the weight for every observed score in every pattern
of observed scores and averaged them across all response patterns
with the same number of responses. For records from students with
one observed score the average weight across the five possible
response patterns is 1.41. For records from students with two
observed scores the average weight on the two scores across all 10
possible response patterns is 2.99. The average of the weights for
records from students with three, four or five observed scores are
3.69, 4.08 and 4.33, respectively. Thus, scores from a student with
five scores will on average have about three times the weight as a
score from a student with just one score. Thus, in the MAR model,
students with few scores are naturally, substantially downweighted.
We believe it is this natural downweighting that resulted in MAR
estimates being robust to violations of MAR in the simulation
studies on missing data and value-added models
[\citet{Wright2004};
\citet{McCaLockMariSeto2005}].

Another potential source for the robustness of teacher effect
estimates is the relatively small scale of changes in student
effects between SEL and MAR.
%Differences in estimated teacher
%effects between the MAR and MNAR models will be driven by
%adjustments student effects make to observed scores and weighting of
%those adjusted scores in the estimation. Using the selection model
%(SEL) rather than MAR changes the estimated student effects and the
%adjusted scores used for estimating teacher effects, but these
%changes are generally very small (two to four percent) relative to
%the variance among classroom average adjusted scores, which is large
%for these data.
%%Even after
%%we control student effects there is substantial variance among the
%classes.
For instance, changes in estimated student effects were only on the
scale of about two to four percent of variance among the classroom
average of the adjusted scores, whereas variation among classrooms
or teachers was large, explaining between 63 and 73 percent of the
variance in the adjusted scores from SEL, depending on the grade.
%The average squared difference between the estimated
%student effects from the SEL and MAR models is only about one
%percent of the overall variance in the student effects from the MAR
%model and the variance in the classroom averages of these
%differences in the estimated student effects is less than one
%percent of the variance of the classroom averages of the adjusted
%scores.
%Hence, the impact of the selection model on estimated
%student effects and adjusted scores is very small (two percent or
%less) relative to the strong classroom heterogeneity in adjusted
%scores. Because SEL can allow for some adjustment of scores for
%students with a single observed score and the MAR model does not,
%this could potentially have an impact on scores but as described
%above, scores from students with a single score are greatly
%downweighted so differences in the adjustments for these students
%have little impact on estimated teacher effects.
%% number in previous paragraph are from deltadiff01.r %%

By allowing the means to differ by response patterns, the pattern
mixture model adjusts student scores differentially by their pattern
of responses. However, as discussed above, the estimated student
effects mostly offset these adjustments, so that the final
adjustments to student scores are similar between the MAR and PMIX.
Scores from students with a single score receive a larger adjustment
with PMIX than MAR, but the downweighting of these scores dampens
the effect of differential adjustments for these students on
estimated teacher effects.

Another factor that potentially contributed to the robustness of
teacher effects to assumptions about missing data is the fact that
scores are observed for the years students are assigned to the
teachers of interest but missing scores in other years. If
observed, the missing data primarily would be used to adjust the
scores from years when students are taught by the teachers of
interest. Our missing data problem is analogous to missing
covariates in linear regression. It is not analogous to trying to
impute values used to estimate group means. In our experience,
estimates of group means from an incomplete sample tend to be more
sensitive to assumptions about missing data than are estimates of
regression coefficients from data with missing covariate values. We
may be seeing a similar phenomenon here.
%In some longitudinal
%student test score databases, scores for students who are not in a
%teacher's class for the entire year are deleted from the data
%because teachers are held accountable only for students who were in
%their class all or nearly all year. In these situations, there
%would be more records with student-teacher links but missing student
%achievement data and estimates might be more sensitive to
%assumptions about missing data.
%%\edit{JRL: CONSIDER DELETING REMAINDER OF PARAGRAPH}
%%I agree but it was in response to a reviewer -- probably won't go
%back to the reviewer can we delete?
%Also, in some instances students linked to a teacher may only
%complete some test items or testing in some subjects, and in these
%instances including the item scores or subsets of test scores could
%reduce the amount of missing data or information further reducing
%the sensitivity of results to MAR assumptions when modeling these
%incomplete data. In these cases, MNAR models like those in

The estimated teacher effects may also be robust to our MNAR models
because these models are relatively modest deviations from MAR.
Although our selection models allowed the probability of observing a
score to depend on each student's general level of achievement, it
did not allow the probability of observing a score to be related
directly to the student's unique level of achievement in a given
year. Such a model might yield greater changes to student effect
estimates and subsequently to estimated teacher effects. The pattern
mixture model did not place such restrictions on selection; however,
it did assume that both the teacher effects and the out-year weights
on those effects did not depend on response patterns. Again, more
flexible models for these parameters might make teacher effects more
sensitive to the model. However, our model specifications are
well aligned with our expectations about missing data mechanisms.
%% In particular, the literature on mobility and student outcomes
%suggests
%% that students with many personal risk factors for low achievement
%and low
%% student effects are more likely to be mobile and hence more likely
%to have
%% incomplete test score records. Moreover, mobility in this population
%is
%% unlikely to be related to the test scores from any particular year.
Also, studies of the heterogeneity of teacher effects as a function
of student achievement have found that such interactions are very
small (explaining three to four percent of the variance in teacher
effects for elementary school teachers [\citet{LockMcCa2009}]).
Hence, it is reasonable to assume that teacher effects would not
differ by response pattern even if response patterns are highly
correlated with achievement.
%Also, we fit a variant of the PMIX
%model in which the student residuals were multivariate normal
%vectors with unspecified covariance matrices
%pattern and again obtained nearly identical estimates of our teacher
%effects.

Downweighting data from students with incomplete data when
calculating the posterior means of teacher effects may be beneficial
beyond making the models robust to assumptions about missing data. A
primary concern with using longitudinal student achievement data to
estimate teacher effects is the potential confounding of estimated
teacher effects with differences in student inputs among classes due
to purposive assignment of students to classes
[\citet{LockMcCa2007EJS}].
%by school personnel, possibly at
%the request of families, or by families in their choice of residence
%or schools.
Although, under relatively unrestrictive assumptions, such biases
can be negated by large numbers of observed test scores on students,
with few tests, the confounding of estimated teacher effects can be
significant [\citet{LockMcCa2007EJS}]. Incomplete data result in
some students with very limited numbers of test scores and the
potential to confound their background with estimated teacher
effects. By downweighting the contributions of these students to
teacher effects, the model mitigates the potential for bias from
purposive assignment, provided some students have a significant
number of observed scores.

We demonstrated that MNAR models can be adapted to the sequential
multi-membership models used to estimate teacher effects from
longitudinal student achievement data, but in our analysis little was
gained from fitting the more complex models. Fitting MNAR models
might still be beneficial in VA modeling applications where the
variability in teacher effects is smaller so that differences in the
estimates of student effects could have a greater impact on
inferences about teachers or where more students are missing scores
in the years they are taught by teachers of interest. A potential
advantage to our selection model is that it provided a means of
controlling for a~student-level covariate (the number of observed
test scores) by modeling the relationship between that variable and
the latent student effect rather than including it in the mean
structure as fixed effect (as was done by PMIX). This approach to
controlling for a covariate might be used more broadly to control
for other variables, such as participation in special programs or
family inputs to education, without introducing the potential for
overcorrecting that has been identified as a possible source of bias
when covariates are included as fixed effects but teacher effects
are random.

 \newpage

\begin{appendix}
\section*{Appendix}\label{appm}
%s5.1 ###
\subsection{Posterior means and standard deviations for parameters of
MAR, SEL and PMIX models}

\mbox{}

%t2 ###
\begin{table}[h]
\tabcolsep=0pt
\caption{Posterior means and standard deviations for yearly means
from pattern mixture model by response pattern. Pattern 25 combines
students with seven rare response patterns}\label{tab:mus1}
\begin{tabular*}{\textwidth}{@{\extracolsep{\fill}}lcccccccccc@{}}
\hline
& \multicolumn{2}{c}{$\bolds{\mu_1}$} & \multicolumn{2}{c}{$\bolds{\mu_2}$}
&\multicolumn{2}{c}{$\bolds{\mu_3}$} & \multicolumn{2}{c}{$\bolds{\mu_4}$} &
\multicolumn{2}{c@{}}{$\bolds{\mu_5}$} \\[-5pt]
& \multicolumn{2}{c}{\hrulefill} & \multicolumn{2}{c}{\hrulefill}
&\multicolumn{2}{c}{\hrulefill} & \multicolumn{2}{c}{\hrulefill} &
\multicolumn{2}{c@{}}{\hrulefill}\\
%(lr){8-9}\cmidrule(lr){10-11}
& \multicolumn{2}{c}{\textbf{Posterior}} & \multicolumn{2}{c}{\textbf{Posterior}}
&\multicolumn{2}{c}{\textbf{Posterior}} & \multicolumn{2}{c}{\textbf{Posterior}} &
\multicolumn{2}{c@{}}{\textbf{Posterior}}\\[-5pt]
& \multicolumn{2}{c}{\hrulefill} & \multicolumn{2}{c}{\hrulefill}
&\multicolumn{2}{c}{\hrulefill} & \multicolumn{2}{c}{\hrulefill} &
\multicolumn{2}{c@{}}{\hrulefill}\\
\textbf{Pattern} & \textbf{Mean} & \textbf{SD}& \textbf{Mean} & \textbf{SD}& \textbf{Mean} & \textbf{SD}& \textbf{Mean} &
\textbf{SD}& \textbf{Mean} & \textbf{SD} \\
\hline
\hphantom{2}1 & 3.71 & 0.03 & 4.32 & 0.04 & 5.11 & 0.03 & 5.70 & 0.03 & 6.34 &
0.03\\
\hphantom{2}2 & NA & NA & 4.27 & 0.05 & 5.00 & 0.04 & 5.52 & 0.04 & 6.16 & 0.04\\
\hphantom{2}3 & 3.67 & 0.06 & NA & NA & 5.06 & 0.06 & 5.69 & 0.05 & 6.35 & 0.05\\
\hphantom{2}4 & NA & NA & NA & NA & 4.98 & 0.04 & 5.53 & 0.04 & 6.17 & 0.04\\
\hphantom{2}5 & 3.57 & 0.07 & 4.24 & 0.07 & NA & NA & 5.56 & 0.06 & 6.26 & 0.06\\
\hphantom{2}6 & NA & NA & 4.07 & 0.11 & NA & NA & 5.39 & 0.10 & 6.00 & 0.09\\
\hphantom{2}7 & 3.51 & 0.14 & NA & NA & NA & NA & 5.64 & 0.14 & 6.21 & 0.12\\
\hphantom{2}8 & NA & NA & NA & NA & NA & NA & 5.52 & 0.04 & 6.22 & 0.04\\
\hphantom{2}9 & NA & NA & NA & NA & NA & NA & NA & NA & 6.15 & 0.06\\
10 & 3.48 & 0.06 & 3.97 & 0.05 & 4.75 & 0.05 & 5.30 & 0.05 & NA & NA\\
11 & NA & NA & 3.91 & 0.06 & 4.55 & 0.06 & 5.09 & 0.06 & NA & NA\\
12 & 3.04 & 0.09 & NA & NA & 4.11 & 0.08 & 4.70 & 0.07 & NA & NA\\
13 & NA & NA & NA & NA & 4.32 & 0.05 & 4.90 & 0.05 & NA & NA\\
14 & 3.21 & 0.13 & 3.79 & 0.13 & NA & NA & 4.93 & 0.11 & NA & NA\\
15 & NA & NA & 3.62 & 0.17 & NA & NA & 4.80 & 0.14 & NA & NA\\
16 & 3.30 & 0.18 & NA & NA & NA & NA & 4.96 & 0.17 & NA & NA\\
17 & NA & NA & NA & NA & NA & NA & 5.02 & 0.06 & NA & NA\\
18 & 3.45 & 0.05 & 4.04 & 0.05 & 4.66 & 0.05 & NA & NA & NA & NA\\
19 & NA & NA & 3.95 & 0.07 & 4.63 & 0.07 & NA & NA & NA & NA\\
20 & 3.28 & 0.09 & NA & NA & 4.48 & 0.10 & NA & NA & NA & NA\\
21 & NA & NA & NA & NA & 4.67 & 0.06 & NA & NA & NA & NA\\
22 & 3.22 & 0.04 & 3.67 & 0.04 & NA & NA & NA & NA & NA & NA\\
23 & NA & NA & 3.92 & 0.05 & NA & NA & NA & NA & NA & NA\\
24 & 3.03 & 0.04 & NA & NA & NA & NA & NA & NA & NA & NA\\
25 & 3.28 & 0.19 & 3.96 & 0.18 & 4.59 & 0.12 & 5.82 & 0.11 & NA & NA\\
\hline
\end{tabular*}
\end{table}
\newpage

%t3 ###
\begin{table}[h]
\tabcolsep=0pt
\caption{Posterior means and standard deviations for student
residual standard deviations from pattern mixture model by response
pattern. Pattern 25 combines students with seven rare response
patterns}\label{tab:rsds1}
\begin{tabular*}{\textwidth}{@{\extracolsep{\fill}}lcccccccccc@{}}
\hline
& \multicolumn{2}{c}{$\bolds{\sigma_1}$} & \multicolumn{2}{c}{$\bolds{\sigma_2}$}
&\multicolumn{2}{c}{$\bolds{\sigma_3}$} & \multicolumn{2}{c}{$\bolds{\sigma_4}$} &
\multicolumn{2}{c@{}}{$\bolds{\sigma_5}$} \\[-5pt]
& \multicolumn{2}{c}{\hrulefill} & \multicolumn{2}{c}{\hrulefill}
&\multicolumn{2}{c}{\hrulefill} & \multicolumn{2}{c}{\hrulefill} &
\multicolumn{2}{c@{}}{\hrulefill}\\
%(lr){8-9}\cmidrule(lr){10-11}
& \multicolumn{2}{c}{\textbf{Posterior}} & \multicolumn{2}{c}{\textbf{Posterior}}
&\multicolumn{2}{c}{\textbf{Posterior}} & \multicolumn{2}{c}{\textbf{Posterior}} &
\multicolumn{2}{c@{}}{\textbf{Posterior}}\\[-5pt]
& \multicolumn{2}{c}{\hrulefill} & \multicolumn{2}{c}{\hrulefill}
&\multicolumn{2}{c}{\hrulefill} & \multicolumn{2}{c}{\hrulefill} &
\multicolumn{2}{c@{}}{\hrulefill}\\
\textbf{Pattern} & \textbf{Mean} & \textbf{SD}& \textbf{Mean} & \textbf{SD}& \textbf{Mean} & \textbf{SD}& \textbf{Mean} &
\textbf{SD}& \textbf{Mean} & \textbf{SD} \\
\hline
\hphantom{2}1 & 0.57 & 0.01 & 0.44 & 0.01 & 0.41 & 0.01 & 0.35 & 0.01 & 0.38 &
0.01\\
\hphantom{2}2 & NA & NA & 0.50 & 0.02 & 0.44 & 0.02 & 0.35 & 0.02 & 0.36 & 0.02\\
\hphantom{2}3 & 0.52 & 0.03 & NA & NA & 0.38 & 0.03 & 0.34 & 0.03 & 0.43 & 0.03\\
\hphantom{2}4 & NA & NA & NA & NA & 0.46 & 0.02 & 0.35 & 0.02 & 0.35 & 0.02\\
\hphantom{2}5 & 0.48 & 0.04 & 0.53 & 0.04 & NA & NA & 0.37 & 0.03 & 0.36 & 0.04\\
\hphantom{2}6 & NA & NA & 0.59 & 0.08 & NA & NA & 0.53 & 0.07 & 0.41 & 0.07\\
\hphantom{2}7 & 0.55 & 0.10 & NA & NA & NA & NA & 0.57 & 0.09 & 0.32 & 0.10\\
\hphantom{2}8 & NA & NA & NA & NA & NA & NA & 0.42 & 0.03 & 0.28 & 0.04\\
\hphantom{2}9 & NA & NA & NA & NA & NA & NA & NA & NA & 0.49 & 0.04\\
10 & 0.60 & 0.03 & 0.46 & 0.02 & 0.41 & 0.02 & 0.44 & 0.02 & NA & NA\\
11 & NA & NA & 0.46 & 0.03 & 0.40 & 0.04 & 0.55 & 0.03 & NA & NA\\
12 & 0.62 & 0.06 & NA & NA & 0.52 & 0.05 & 0.38 & 0.05 & NA & NA\\
13 & NA & NA & NA & NA & 0.48 & 0.03 & 0.39 & 0.04 & NA & NA\\
14 & 0.62 & 0.08 & 0.48 & 0.08 & NA & NA & 0.38 & 0.09 & NA & NA\\
15 & NA & NA & 0.62 & 0.11 & NA & NA & 0.27 & 0.15 & NA & NA\\
16 & 0.51 & 0.15 & NA & NA & NA & NA & 0.39 & 0.17 & NA & NA\\
17 & NA & NA & NA & NA & NA & NA & 0.85 & 0.04 & NA & NA\\
18 & 0.53 & 0.03 & 0.45 & 0.03 & 0.58 & 0.03 & NA & NA & NA & NA\\
19 & NA & NA & 0.48 & 0.06 & 0.56 & 0.05 & NA & NA & NA & NA\\
20 & 0.36 & 0.10 & NA & NA & 0.66 & 0.07 & NA & NA & NA & NA\\
21 & NA & NA & NA & NA & 0.96 & 0.03 & NA & NA & NA & NA\\
22 & 0.48 & 0.03 & 0.54 & 0.02 & NA & NA & NA & NA & NA & NA\\
23 & NA & NA & 0.84 & 0.03 & NA & NA & NA & NA & NA & NA\\
24 & 0.98 & 0.01 & NA & NA & NA & NA & NA & NA & NA & NA\\
25 & 0.67 & 0.12 & 0.65 & 0.12 & 0.29 & 0.09 & 0.24 & 0.10 & NA & NA\\
\hline
\end{tabular*}
\end{table}
\newpage

%t4 ###
\begin{table}[h!]
\tabcolsep=0pt
\tablewidth=300pt
\caption{Posterior means and standard deviations for standard
deviation of student effects, by response pattern, standard
deviation of teacher effects and prior teacher effect weights, which
are constant across response pattern for the pattern mixture model.
Response patterns 9, 17, 21, 23 and 24 involve a single observation
so all the student variance is modeled by the residual variance and
there are no additional student effects or standard error of student
effects estimated for these patterns}\label{tab:oparm1}
\begin{tabular*}{300pt}{@{\extracolsep{\fill}}lcc@{}}
\hline
&\multicolumn{2}{c@{}}{\textbf{Posterior}}\\[-5pt]
&\multicolumn{2}{c@{}}{\hrulefill}\\
\textbf{Parameter}&\textbf{Mean}&\textbf{Std. dev.}\\
\hline
$\nu$, Pattern 1 & 0.62 & 0.01\\
$\nu$, Pattern 2 & 0.63 & 0.02\\
$\nu$, Pattern 3 & 0.63 & 0.03\\
$\nu$, Pattern 4 & 0.60 & 0.02\\
$\nu$, Pattern 5 & 0.47 & 0.04\\
$\nu$, Pattern 6 & 0.44 & 0.07\\
$\nu$, Pattern 7 & 0.66 & 0.09\\
$\nu$, Pattern 8 & 0.60 & 0.03\\
$\nu$, Pattern 10 & 0.73 & 0.03\\
$\nu$, Pattern 11 & 0.70 & 0.04\\
$\nu$, Pattern 12 & 0.69 & 0.06\\
$\nu$, Pattern 13 & 0.71 & 0.03\\
$\nu$, Pattern 14 & 0.68 & 0.08\\
$\nu$, Pattern 15 & 0.67 & 0.11\\
$\nu$, Pattern 16 & 0.80 & 0.14\\
$\nu$, Pattern 18 & 0.71 & 0.03\\
$\nu$, Pattern 19 & 0.74 & 0.05\\
$\nu$, Pattern 20 & 0.70 & 0.07\\
$\nu$, Pattern 22 & 0.66 & 0.02\\
$\nu$, Pattern 25 & 0.52 & 0.09\\
$\tau_1$ & 0.63 & 0.03\\
$\tau_2$ & 0.55 & 0.03\\
$\tau_3$ & 0.51 & 0.02\\
$\tau_4$ & 0.41 & 0.02\\
$\tau_5$ & 0.43 & 0.02\\
$\alpha_{21}$ & 0.14 & 0.02\\
$\alpha_{31}$ & 0.12 & 0.02\\
$\alpha_{32}$ & 0.19 & 0.02\\
$\alpha_{41}$ & 0.08 & 0.02\\
$\alpha_{42}$ & 0.10 & 0.02\\
$\alpha_{43}$ & 0.11 & 0.02\\
$\alpha_{51}$ & 0.08 & 0.02\\
$\alpha_{52}$ & 0.14 & 0.02\\
$\alpha_{53}$ & 0.07 & 0.02\\
$\alpha_{54}$ & 0.32 & 0.03\\
\hline
\end{tabular*}
\end{table}

%s5.2 ###
\subsection{Results for the alternative selection model}
\begin{figure}

\includegraphics{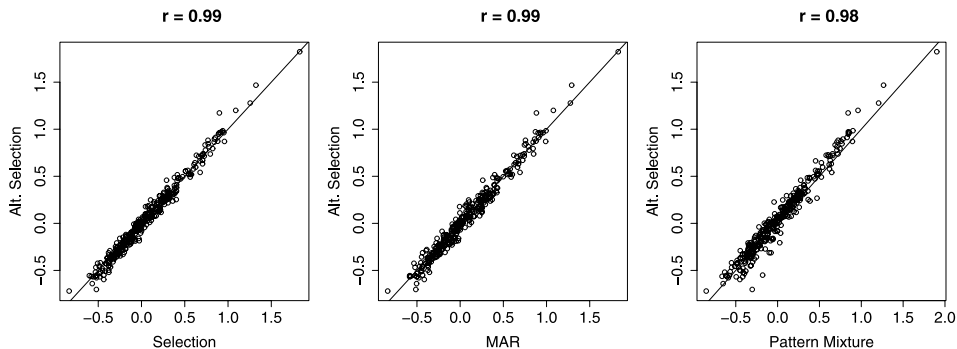}

\caption{Scatter plots of posterior means for fourth
grade teacher effects from selection, pattern mixture and MAR models
versus those from the alternative selection model.}
\label{fig:thetas2}
\end{figure}

Figure \ref{fig:thetas2} compares the estimated teacher effects from
the alternative selection and other models. The correlation between
the estimated teacher effects from this alternative selection model
and those from SEL were 0.97, 0.98, 0.99, 0.99 and 1.00, for grades
one to five, respectively. As shown in Figure \ref{fig:thetas2},
the estimated fourth grade teacher effects from the new model are
not only highly correlated with those from SEL, they are nearly
identical to those from all the other models.   Other grades are
similar.

%
%f6 ###
%f7 ###
\begin{figure}[b]

\includegraphics{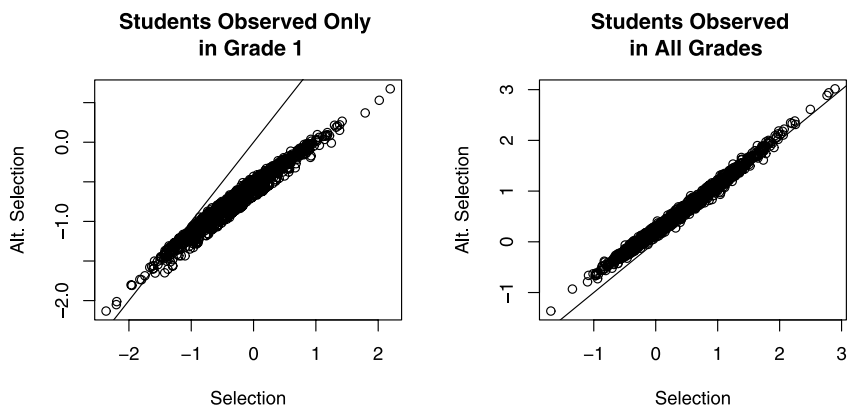}

\caption{Scatter plots of posterior means from the two
alternative selection models for effects of students with selected
response patterns.} \label{fig:deltas2}
\end{figure}

The two alternative selection models do, however, yield somewhat
different estimates of the individual student effects. The
differences were most pronounced for students observed only in grade
one in which the alternative selection model tended to shift the
distribution of these students toward lower levels of achievement
(left panel of Figure \ref{fig:deltas2}). However, differences even
exist for students observed at every grade (right panel of
Figure \ref{fig:deltas2}). Again, these differences are
sufficiently small or the students are sufficiently downweighted so
that they do not result in notable changes to the estimated teacher
effects.
\end{appendix}

\section*{Acknowledgment}
We thank
Harold C. Doran for providing us the data used in the analyses presented
in this article.

\begin{supplement}%[id=suppA]
\stitle{Student achievement data and WinBUGS code\\}
\slink[doi,text=10.1214/10-\break AOAS405SUPP]{10.1214/10-AOAS405SUPP}
\slink[url]{http://lib.stat.cmu.edu/aoas/405/supplement.zip}
\sdatatype{.zip}
\sdescription{The file SHAR generates six files:
\begin{enumerate}[6.]
\item readme.
\item AOAS405\_McCaffrey\_Lockwood\_MNAR.csv
contains the 1998--2002 student achievement data with student and teacher
identifiers used to estimate teacher effects using selection and
pattern mixture models. The comma delimited file contains four
variables:
\begin{enumerate}[(b)]
\item[(a)] stuid -- student ID that is common among records from the same
teacher;
\item[(b)] tchid -- teacher ID that is common among students in
the teacher's class during a year;
\item[(c)] year -- indicator of year of
data takes on values 0--4 (grade level equals year${}+{}$1);
\item[(d)] Y -- student's district mathematics test score for year
rescaled by subtracting 400 and dividing by 40.
\end{enumerate}
\item AOAS405\_McCaffrey\_Lockwood\_MAR-model.txt -- Annotated WinBUGS
code used for fitting Model (\ref{eq:mar1})
assuming data are missing at random (MAR).
\item AOAS405\_McCaffrey\_Lockwood\_sel-model.txt -- Annotated WinBUGS
co\-de used for fitting Model (\ref{eq:mar1})
with assumption 1 for missing data.
\item AOAS405\_McCaffrey\_Lockwood\_sel2-model.txt -- Annotated
WinBUGS code used for fitting Model (\ref{eq:mar1})
with assumption 1b for missing data.
\item AOAS405\_McCaffrey\_Lockwood\_patmix-model.txt -- Annotated~Win-\break BUGS code used for fitting the
pattern mixture Model (\ref{eq:patmix1}).
\end{enumerate}
}
\end{supplement}

%suskaldyti doi

\printaddresses

\end{document}